\documentclass[reprint,secnumarabic, graphics,floatfix,nofootinbib,
tightenlines,nobibnotes,aps,prl,superscriptaddress]{revtex4-1}

\usepackage{graphicx}
\graphicspath{}
\usepackage{dcolumn}
\usepackage{bm}
\usepackage{mathtools}
\usepackage{physics}
\usepackage{tikz}
\usetikzlibrary{arrows}
\tikzset{
    level/.style = {
        ultra thick,
        black,
    },
    connect/.style = {
        dashed,
        black
    },
    label/.style = {
        text width=2cmz`
    },
     trans/.style = {
     	thick,<->,shorten >=2pt,shorten <=2pt,>=stealth
	}
}
\usepackage{hyperref}

\begin{document}


\title{Interaction Spectroscopy of a Two-component Mott Insulator}

\author{Jesse Amato-Grill}
\affiliation{Research Laboratory of Electronics, MIT-Harvard Center for Ultracold Atoms, and
Department of Physics, Massachusetts Institute of Technology, Cambridge, Massachusetts 02139, USA} 
 \affiliation{Department of Physics, Harvard University, Cambridge, Massachusetts 02138, USA}
\author{Niklas Jepsen}
 \affiliation{Research Laboratory of Electronics, MIT-Harvard Center for Ultracold Atoms, and
 Department of Physics, Massachusetts Institute of Technology, Cambridge, Massachusetts 02139, USA}
 \author{Ivana Dimitrova}
  \affiliation{Research Laboratory of Electronics, MIT-Harvard Center for Ultracold Atoms, and
Department of Physics, Massachusetts Institute of Technology, Cambridge, Massachusetts 02139, USA}
 \author{William Lunden}
  \affiliation{Research Laboratory of Electronics, MIT-Harvard Center for Ultracold Atoms, and
Department of Physics, Massachusetts Institute of Technology, Cambridge, Massachusetts 02139, USA}
 \author{Wolfgang Ketterle}
  \affiliation{Research Laboratory of Electronics, MIT-Harvard Center for Ultracold Atoms, and
Department of Physics, Massachusetts Institute of Technology, Cambridge, Massachusetts 02139, USA}

\date{\today}
\begin{abstract}
We prepare and study a two-component Mott insulator of bosonic atoms with two particles per site. The mapping of this system to a magnetic spin model, and the subsequent study of its quantum phases, require a detailed knowledge of the interaction strengths of the two components. In this work, we use radio frequency (RF) transitions and an on-site interaction blockade for precise, empirical determination of the interaction strengths of different combinations of hyperfine states on a single lattice site. We create a map of the interactions of the lowest two hyperfine states of $^7$Li as a function of magnetic field, including measurements of several Feshbach resonances with unprecedented sensitivity, and we identify promising regions for the realization of magnetic spin models.

\end{abstract}

\maketitle

Ultracold atoms in optical lattices, described by a Hubbard Hamiltonian, are a uniquely accessible platform for the study of quantum magnetism. A dual-component Mott insulator with $n$ atoms per site, in which each of the components stands in for a magnetic spin, implements a spin-$n/2$ Heisenberg model with nearest neighbor interactions \cite{altman}. With first-order tunneling suppressed by on-site interactions, only exchanges between sites that preserve the overall density distribution are possible. This \textit{superexchange} of particles mediates effective spin-spin interactions \cite{duan}, and in analogy to the spin system, the ground state will be determined by ratios of the on-site intra- and inter-species interactions, which can be varied by means of a state-dependent optical lattice \cite{bec4} or by using Feshbach resonances \cite{chin}. The model with a single particle per site, corresponding to spin-1/2, has been studied extensively in many regimes; recent successes include observation of 3D and 2D N\'eel ordering in fermions \cite{hart,mazurenko}, measurements of spin correlations in 2D spin-imbalanced systems \cite{brown} and spin-charge correlations in the presence of hole doping \cite{cheuk}.
\begin{figure}[]
\begin{tikzpicture}[scale=.6]
	\definecolor{mlred}{RGB}{217,83,25}
	\definecolor{mlblue}{RGB}{0,114,189}
	\definecolor{mlredfill}{RGB}{236,169,140}
	\definecolor{mlbluefill}{RGB}{127,184,222}
	
	\draw[level] (0,0) -- node[above] {$\ket{aa}$} (3,0);
	\draw[level] (0,2) -- node[above] {$\frac{1}{\sqrt{2}}(\ket{ab}+\ket{ba})$} (3,2);
	\draw[level] (0,4) -- node[above] {$\ket{bb}$} (3,4);
	
	\draw[connect](3,0) -- (6,-2) (3,2) -- (6,2.5) (3,4) -- (6,5);
	
	\draw[level] (6,-2) -- node[above] {} (9,-2);
	\draw[level] (6,2.5) -- node[above] {} (9,2.5);
	\draw[level] (6,5) -- node[above] {} (9,5);
	
	\draw[trans] (7.5,-2) -- (7.5,2.5) node[midway,right] {$\omega_{aa\rightarrow ab}$};
    	\draw[trans] (7.5,2.5) -- (7.5,5) node[midway,right] {$\omega_{bb\rightarrow ab}$};

	\draw[-,>=stealth](0,-3)-- (3,-3);
	\draw[connect](3,-3)--(6,-3) node[midway,below] {$B$} ;
	\draw [->,>=stealth] (6,-3) -- (9,-3);
	\begin{scope}[shift={(-2,0)}]
		\draw [black,thick,domain=(-3*pi/4):(7*pi/4),xscale=.25,yscale=.75] plot (\x, {-sin(\x r)});
		\draw [draw=mlred, thin, fill=mlred] (.4,-.25) circle [radius=0.2];
		\draw [draw=mlred, thin, fill=mlred] (.4,.25) circle [radius=0.2];
	\end{scope}
	\begin{scope}[shift={(-2,2)}]
		\draw [black,thick,domain=(-3*pi/4):(7*pi/4),xscale=.25,yscale=.75] plot (\x, {-sin(\x r)});
		\draw [draw=mlred, thin, fill=mlred] (.4,-.25) circle [radius=0.2];
		\draw [draw=mlblue, thin, fill=mlblue] (.4,.25) circle [radius=0.2];
	\end{scope}
	\begin{scope}[shift={(-2,4)}]
		\draw [black,thick,domain=(-3*pi/4):(7*pi/4),xscale=.25,yscale=.75] plot (\x, {-sin(\x r)});
		\draw [draw=mlblue, thin, fill=mlblue] (.4,-.25) circle [radius=0.2];
		\draw [draw=mlblue, thin, fill=mlblue] (.4,.25) circle [radius=0.2];
	\end{scope}
\end{tikzpicture}
\caption{The three different combinations of two hyperfine states on a lattice site have interaction energies $U_{aa}$, $U_{ab}$, and $U_{bb}$ which can be tuned via Feshbach resonances. At a given field $B$, the splittings will in general be unequal, giving rise to an interaction blockade: the two possible transitions can be individually addressed by an RF drive with frequency $\omega=\omega_{Zeeman}+\Delta U/\hbar$. }
\label{fig:1}
\end{figure}
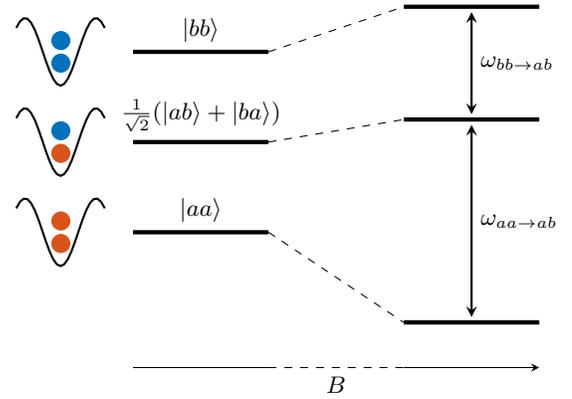

In this work, we focus on a two-component, spin-1 bosonic model implemented using the two lowest hyperfine states of $^7$Li (hereafter $a$ and $b$) in a cubic optical lattice. Integer-spin models remain largely unexplored using cold atoms systems, and yet they are predicted to exhibit many interesting behaviors arising from both magnetic ordering and from beyond-mean-field effects due to topological considerations \cite{afflek,renard,wierschem2}. With two particles per site, the three states of our model, $\ket{+1}\,{=}\,\ket{aa}$, $\ket{0}\,{=}\,(\ket{ab}\,{+}\,\ket{ba})/\sqrt{2}$ and $\ket{-1}\,{=}\,\ket{bb}$, form a spin triplet manifold on each lattice site.\footnote{Hereafter, for compactness, we will refer to these three states as $\ket{aa}$, $\ket{ab}$, and $\ket{bb}$, using the Fock basis notation.} As in the spin-1/2 system, the nearest neighbor interactions arise from superexchange. Because of Feshbach resonances, the three possible configurations have on-site energies that are a function of the applied magnetic field $B$, so that one can tune the configurational energies of different distributions of spins in the lattice (Fig. \ref{fig:1}). Quantum phase transitions are expected in regimes where the nearest-neighbor interactions compete with on-site interactions. Determining the relevant regions of $B$ for an exploration of the spin-1 Hamiltonian therefore requires precise knowledge of the relative strength of the interactions amongst the hyperfine states.

\begin{figure}[]
\includegraphics[width=\linewidth, trim={0 0 0 0},clip]{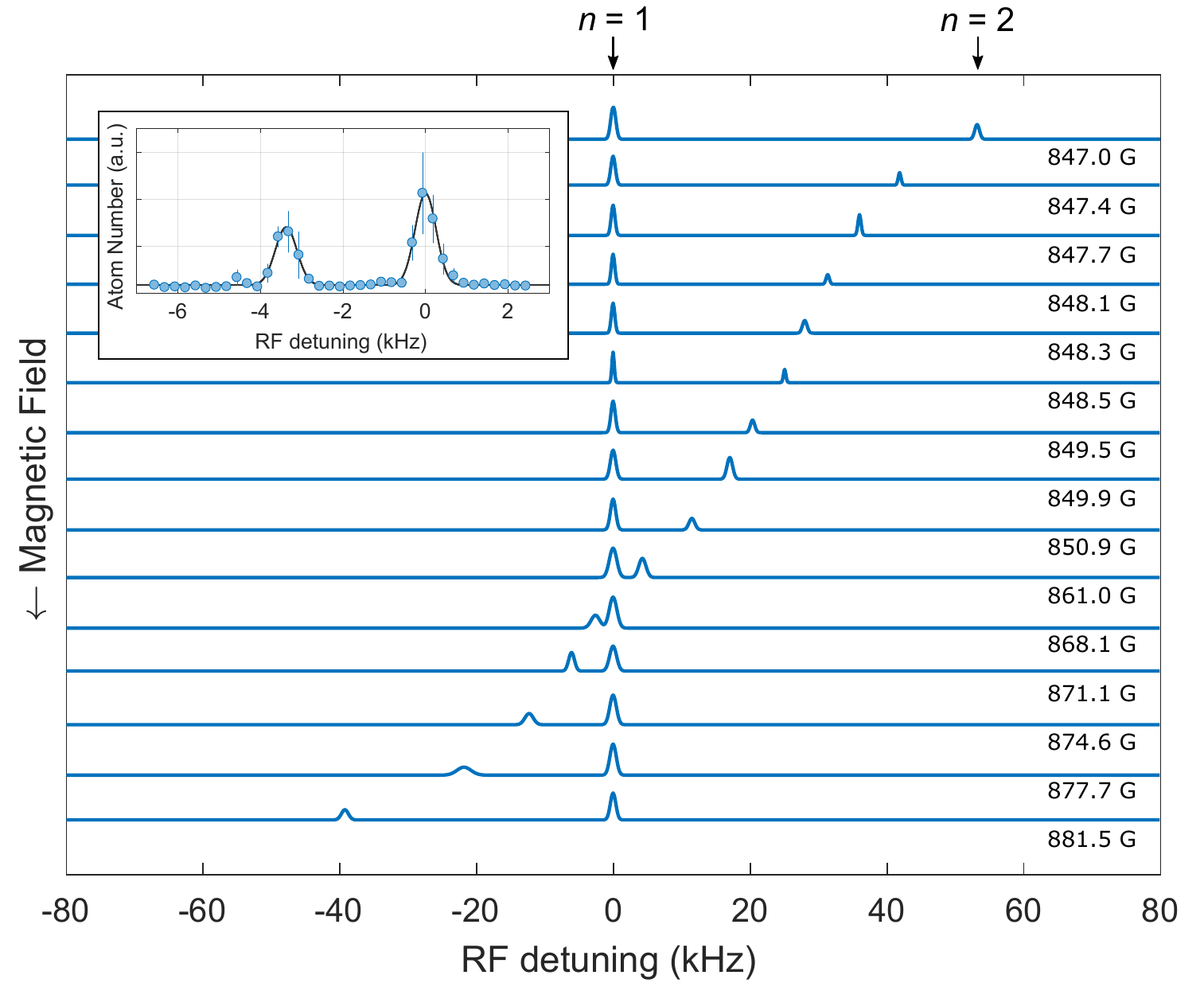}
\caption{Fits to representative spectra of the transition between $b$ and $a$, taken between the two $bb$ Feshbach resonances. The spectra have been plotted so that the Zeeman-shifted peaks of the $n\,{=}\,1$ transitions overlap. The inset shows an example spectrum in which each point is an average of four measurements and the fit is a sum of two Gaussians. The crossing of the frequencies of the two peaks corresponds to $U_{ab}\,{-}\,U_{bb}\,{=}\,0.$}
\label{fig:2}
\end{figure}

RF spectroscopy has been used in the past to measure site occupancy in a single-species Mott insulator of ultracold bosons \cite{campbell}. Here we extend this technique to measure the differential interaction energy of two confined atoms as a function of magnetic field. We begin by preparing an $n\,{=}\,2$ Mott insulator of $^7$Li in a single hyperfine state in a 1064nm cubic optical lattice (our apparatus and BEC production are described elsewhere \cite{dimitrova}). The number of atoms is $1\,{\times}\,10^5$ and the lattice depth is 35 $E_R$ in each dimension. The central $n\,{=}\,2$ plateau, approximately $3\,{\times}\,10^4$ sites, is surrounded by an $n\,{=}\,1$ shell containing a similar number of sites. We pulse the RF drive for 2.9ms and monitor the number of atoms in the other hyperfine state as a function of drive frequency. At the frequency corresponding to the transition of a bare atom, we observe a peak coming from the atoms on $n\,{=}\,1$ sites. We observe a second peak from the $n\,{=}\,2$ atoms, which is shifted by the difference in interaction energy between the initial state ($\ket{aa}$ or $\ket{bb}$) and the final state ($\ket{ab}$). The pulse length corresponds to a $\pi$-pulse for the $n\,{=}\,2$ sites (so that we maximize the signal), which have a Rabi frequency $\sqrt{2}$ greater than that of the $n\,{=}\,1$ sites, due to bosonic enhancement. The interaction blockade \cite{bakr} that arises from unequal interaction energies in the three states means that one may drive the system selectively between $\ket{aa}$ and $\ket{ab}$ (or $\ket{bb}$ and $\ket{ab}$). Thus when we probe the atoms absorptively after an RF pulse with light that is resonant only for $a$ (or $b$), we measure a single flipped atom per $n\,{=}\,2$ site.

The frequency of the $n\,{=}\,1$ peak corresponds to the Zeeman shift and thus to the magnitude of the applied magnetic field (the hyperfine constant and nuclear g-factor for $^7$Li are taken from \cite{inguscio}). The frequency shift of the $n\,{=}\,2$ peak, which may be positive or negative, is a direct measure of the differential two-body interactions. Using this technique, we obtain RF spectra at many selected bias fields from which we derive the two-body interaction splittings $U_{bb}-U_{ab}$ and $U_{ab}-U_{aa}$ (Fig. \ref{fig:2}). The technique works equally well for attractive and repulsive interactions, so long as the system remains in the Mott insulating state.

\begin{figure}[h!]
\includegraphics[width=\linewidth, trim={0 0 0 0},clip]{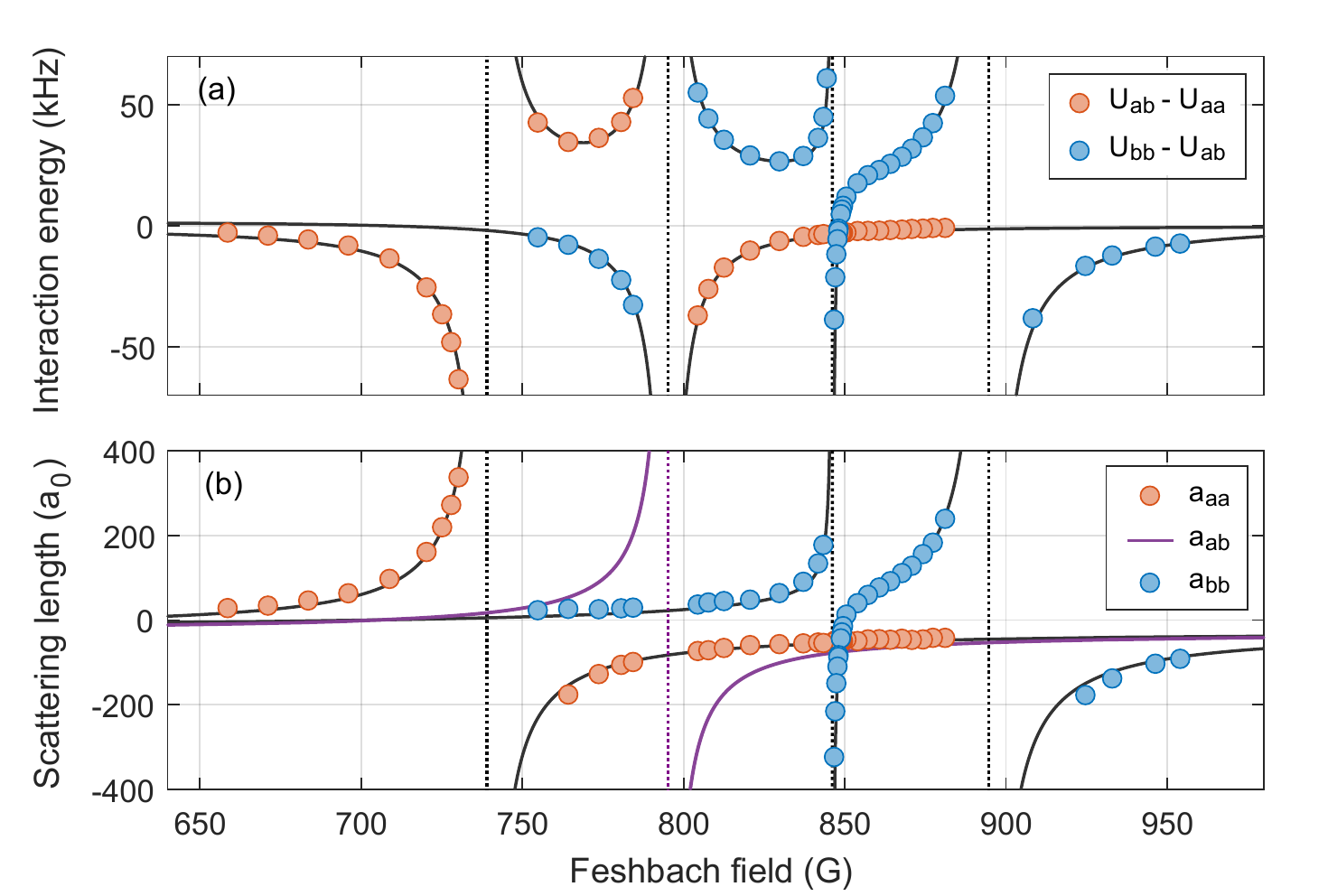}
\caption{ (a) The relative interactions $U_{ab}\,{-}\,U_{aa}$ and $U_{bb}\,{-}\,U_{ab}$ are plotted as a function of magnetic field, measured via RF interaction spectroscopy in 35, 35, 35 $E_R$ optical lattice.  (b) The scattering lengths of the $aa$ and $bb$ interactions are plotted as a function of magnetic field, measured using lattice AM and shown in units of the Bohr radius $a_0$. Also shown as a bold purple line is the $ab$ scattering length, obtained using simultaneous hyperbolic fits to the RF spectroscopy and  lattice AM data sets. The fits are shown as thin black lines.}
\label{fig:3}
\end{figure}

The precision to which we must determine the differential interaction energies on a site as an input to a many-body physics model is fixed by the superexchange rate, which at Mott insulator depths for $^7$Li in a 1064 nm optical lattice ranges from hundreds of hertz to several kilohertz, depending on lattice depth and dimensionality. Here we measure the differential interaction energies as a function of magnetic field to a precision of about 100 Hertz, limited only by the stability of our magnetic field (better than one part in $10^5$ at ~$10^3$ G) and the difference in magnetic moments of the two hyperfine states (approximately 33 kHz/G). As these interaction energies range over many tens of kilohertz, the error bars are too small to see on a full scale plot (Fig. \ref{fig:3}a). This technique is particularly well-suited to atoms with characteristically large interaction energies, such as Li, because the wide separation between singlon and doublon spin-flip resonances permits the use of high Rabi frequencies, which maximizes signal size and decreases sensitivity to magnetic field noise.  

\begin{table*}[t]
\begin{tabular}{c p{.2cm} c p{.2cm} c p{.2cm} c p{.2cm} c p{.2cm} c}
\hline
Channel 	&&  $a_{bg}/a_0$ 	&& $\Delta$ (G) 	&& $B_{res}$ (G)		&& $B_{res}$ (G)		&&  $B_{res}$ (G)	\rule{0pt}{10pt}				\\ 
		&&			&&			&& {\scriptsize RF data only}\par	&& {\scriptsize Combined fit}\par	&&{\scriptsize Previous works}\par						\\\hline\hline
$aa$ 		&& -25.8(1.2)	&& -135.9(6.9)	&& 737.58(10)		&& 738.29(15)		&& 738.2(2)$^{\text{\cite{gross2}}}$	\rule{0pt}{10pt}	\\ 
         		&&			&&			&&				&&				&& 737.8(2)$^{\text{\cite{navon}}}$				\\
         		&&			&&			&&				&&				&& 736.97(7)$^{\text{\cite{pollack}}}$			\\
$ab$ 		&& -29.8(1.3)	&& -90.5(4.0)	&& 794.64(07)		&& 794.59(12)		&&	   								\\ 
$bb$ 		&& -23.0(1.4)	&& -14.9(0.9)	&& 845.42(01)		&& 845.45(02)		&& 844.9(8)$^{\text{\cite{gross2}}}$				\\ 
$bb$ 		&& -23.0(1.4)	&& -172.7(10.0)	&& 893.34(12)		&& 893.84(18)		&& 893.7(4)$^{\text{\cite{gross2}}}$				\\ \hline
\end{tabular}
\caption{Parameters of the Feshbach resonances in the lowest two hyperfine states of $^7$Li, determined with a simultaneous fit to RF interaction and lattice AM spectroscopy data (except where specified otherwise). The reported errors in our measurements are 1$\sigma$ statistical uncertainties in the fit parameters.}\label{table1}
\end{table*}

While RF spectroscopy in a lattice is a powerful and precise tool for characterizing differential interactions, another technique is necessary to measure the absolute interaction energy (i.e. $U_{aa}$ or $ U_{bb}$), for these interactions determine the lattice depth for the transition to the Mott insulator in each hyperfine state. In previous studies, lattice amplitude modulation (AM) has been used to drive singlon-to-doublon conversion in the lowest Hubbard band, and the resonant frequency of this process has been associated with the on-site interaction energy in both repulsive \cite{schori} and attractive \cite{mark} single-component bosonic systems. Here we employ the same technique to map the on-site intra-species interactions of $^7$Li across a broad range of magnetic fields. We modulate the lattice depth by 30\% peak-to-peak along the shallow dimension of a 35, 35, 20 $E_R$ optical lattice and measure the entropy added to the system by adiabatically ramping back to the BEC from the Mott insulator and measuring the recondensed fraction. While we must remain in the Mott insulator in regions of magnetic field where the scattering length is negative in order to prevent the collapse of the atomic cloud, data taken above but close to the transition for either attractive or repulsive interactions display a bias towards higher frequency \cite{schori}. While this systematic bias prevents the collection of data for very small scattering lengths, limiting the accuracy of our determination of the background scattering length in each channel, the locations of the resonances themselves are not significantly affected.

We perform a simultaneous fit to both the RF differential interaction spectroscopy data and the lattice AM spectroscopy data in order to determine the inter-species interaction energy $U_{ab}$, and to extract parameters of the Feshbach resonances we detect (see Fig. \ref{fig:3}b). We take the form of the scattering length in each state to be a hyperbola
\begin{equation}\label{hyperbola}
	a_s=a_{bg}\Big(1-\sum_i\frac{\Delta^{(i)}}{B-B_{res}^{(i)}}\Big)
\end{equation}
with a background scattering length $a_{bg}$, width $\Delta^{(i)}$, and resonance location $B_{res}^{(i)}$. In order to provide a useful comparison with existing literature, we calculate the scattering lengths
\begin{equation}\label{onsite}
	\frac{1}{a_s}=\frac{4\pi\hbar^2}{m}\frac{1}{U}\int|\psi(r)| ^4d^3r
\end{equation}
where $U$ is the on-site interaction energy we measure and $\psi(r)$is the calculated Wannier wavefunction on a site, given calibration of the lattice depth by inter-band parametric excitation. This approximation for $\psi$ systematically biases the scattering lengths towards lower values because interactions modify the actual two-particle wavefunction by admixing higher bands, but the correction is not significant as long as $|a_s|/a_{HO}\ll V^{1/4}/\sqrt{2}\pi$ where $a_{HO}$ is the harmonic oscillator length on a site and $V$ is the lattice depth in recoil units \cite{buchler}. For the moderate scattering lengths considered here, we remain more than one order of magnitude below this threshold. The complete Feshbach resonance spectrum for the lowest two hyperfine states of $^7$Li  is plotted in (Fig. \ref{fig:3}b) and the parameters of the resonances can be found in Table \ref{table1}.

We find a single resonance in the lowest hyperfine state, whose position is in good agreement with the most recent measurements made using RF-spectroscopy of molecular binding energies \cite{gross2}, although previous measurements made using the modification of in-trap condensate size due to the mean field energy of interactions have reported the resonance at slightly lower magnetic fields \cite{pollack, pollack2, navon}. We also find a single resonance between the $a$ and $b$ states, previously unmeasured, whose parameters are particularly relevant for studies of two-component systems. Of interest is also the double resonance in the $b$ state, studied previously using three-body atom loss \cite{gross, shotan} and also RF-spectroscopy of molecular binding energies \cite{gross2}. Compared to previous techniques, RF interaction spectroscopy allows for the exploration of the inter-species resonance. The method is as precise as RF spectroscopy of  molecular binding energies, but does not require knowledge of the molecular potential in order to extract the scattering lengths.

\begin{figure}[h!]
\includegraphics[width=\linewidth, trim={0 0 0 0},clip]{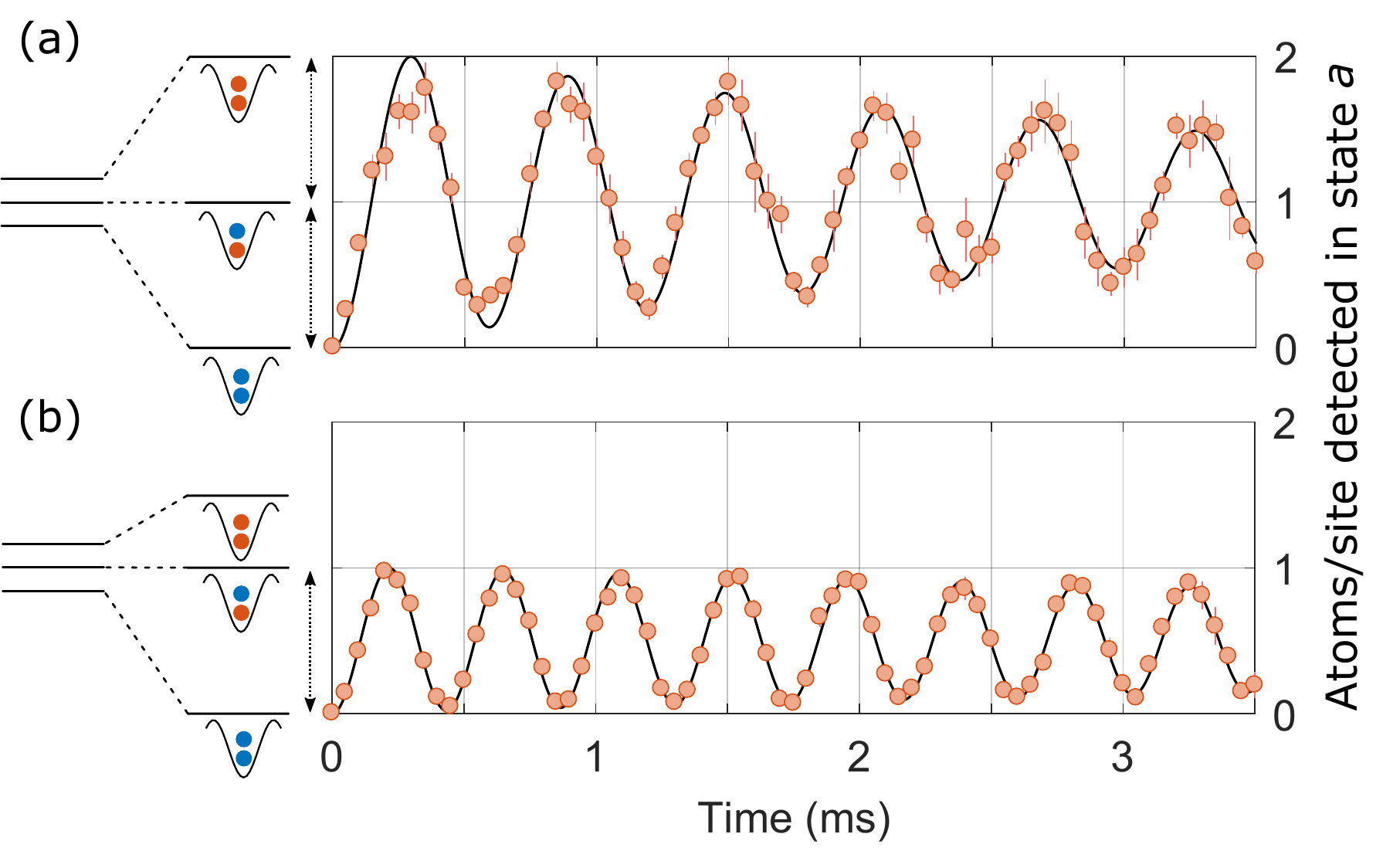}
\caption{Rabi oscillations of doublons (a) exactly at and (b) far away from the magnetic field at which the interactions are degenerate, as in (\ref{eq:dgp}). The measured atom number is normalized to the total number of $n\,{=}\,2$ sites. The system is initially prepared in state $b$ and we detect the total number of atoms in state $a$ after applying a resonant RF drive. Decaying sinusoidal fits determine the Rabi frequencies to be 1.68(4) kHz and 2.30(1) kHz respectively, consistent with the expected $\sqrt{2}$ ratio in Rabi frequency between resonant three-level and two-level systems. The time constant for the decoherence away from degeneracy is 10.2 ms. The oscillations in (a) seem to decay faster because this data was taken at the closest magnetic field to the degeneracy point which is permitted by the present resolution limit of our magnetic field setpoint, so that we ultimately observed beating between two nearly-equal but off-resonant Rabi frequencies.}
\label{fig:4}
\end{figure}

The region between the two $bb$ resonances provides an opportunity to vary the interaction energy of a $\ket{bb}$ site with respect to that of an $\ket{aa}$ or $\ket{ab}$ while keeping the latter two relatively constant. We even find a point, near 849 G, where 
\begin{equation}\label{eq:dgp}
2U_{ab}=U_{aa}+U_{bb}
\end{equation} 
so that the transitions from $\ket{aa}$  to $\ket{ab}$ and $\ket{ab}$ to $\ket{bb}$ occur at the same frequency. The interaction blockade vanishes and it is possible to rotate around the entire spin-1 Bloch sphere. This point is characterized by Rabi oscillations with twice the amplitude (i.e. full contrast on all doubly occupied sites) but $\sqrt{2}$ lower Rabi frequency than when the two transition frequencies are different and the RF drive is resonant with only one of them (Fig. \ref{fig:4}). Moreover, for two sites connected by tunneling, the degeneracy condition (\ref{eq:dgp}) also means that the superexchange process corresponding to $\ket{+1}_L\ket{-1}_R \Rightarrow\ket{0}_L\ket{0}_R$ becomes resonant. This degeneracy implements a special point in the spin-1 Heisenberg model: it is the point where the spin-spin interactions are isotropic, and where the on-site anisotropy, which biases the system towards local pairing, completely vanishes.

As much as RF interaction spectroscopy in a lattice enables precise measurements of scattering lengths, it is also a tool for state preparation and diagnostics. Starting with an $n\,{=}\,2$ Mott insulator in a single hyperfine state, one can prepare the fully paired state $\ket{ab}$ on every doubly-occupied lattice site by means of a $\pi$-pulse or a Landau-Zener sweep. Figure \ref{fig:4}b demonstrates coherent preparation of this fully paired state, or Spin Mott state, which has a large gap and is a promising starting point for adiabatic state preparation, in analogy to the band insulator in fermions \cite{bec4}. Full diagnostics of doubly occupied sites can be realized using transitions to a third hyperfine state, by selectively measuring the number of doublons in different configurations of hyperfine states. We therefore expect this method to prove useful in studies of strongly interacting multi-component cold atoms systems, especially when the relative interaction strengths can be modified with Feshbach resonances or with a spin-dependent lattice. 

In conclusion, we have developed a technique to precisely determine the relative on-site interaction energies of different configurations of two-component bosons in an optical lattice, and have demonstrated that technique using $^7$Li over a broad range of magnetic fields. We have improved the precision parameters of three previously observed Feshbach resonances, and report the first observation of an inter-species Feshbach resonance in $^7$Li. The identification of a magnetic field which provides for a degeneracy of differential interactions paves the way for future investigation of a spin-1 bosonic system, which is predicted to include spin-ordered phases such an XY-antiferromagnet, a Z-paramagnet, and topologically protected phases such as the Haldane phase.

\begin{acknowledgments}
We would like to thank Junru Li and Wenlan Chen for critical readings of this manuscript. We acknowledge support from the NSF through the Center for Ultracold Atoms and award 1506369, from ARO-MURI Non-equilibrium Many-body Dynamics (Grant No. W911NF-14-1-0003), from AFOSR-MURI Quantum Phases of Matter (Grant No. FA9550-14-1- 0035), from ONR (Grant. No. N00014-17-1-2253) and a Vannevar-Bush Faculty Fellowship. J.A-G. acknowledges support by the National Science Foundation Graduate Research Fellowship under Grant No. DGE 1144152. 
\end{acknowledgments}

\bibliography{bib}

\begin{thebibliography}{24}%
\makeatletter
\providecommand \@ifxundefined [1]{%
 \@ifx{#1\undefined}
}%
\providecommand \@ifnum [1]{%
 \ifnum #1\expandafter \@firstoftwo
 \else \expandafter \@secondoftwo
 \fi
}%
\providecommand \@ifx [1]{%
 \ifx #1\expandafter \@firstoftwo
 \else \expandafter \@secondoftwo
 \fi
}%
\providecommand \natexlab [1]{#1}%
\providecommand \enquote  [1]{``#1''}%
\providecommand \bibnamefont  [1]{#1}%
\providecommand \bibfnamefont [1]{#1}%
\providecommand \citenamefont [1]{#1}%
\providecommand \href@noop [0]{\@secondoftwo}%
\providecommand \href [0]{\begingroup \@sanitize@url \@href}%
\providecommand \@href[1]{\@@startlink{#1}\@@href}%
\providecommand \@@href[1]{\endgroup#1\@@endlink}%
\providecommand \@sanitize@url [0]{\catcode `\\12\catcode `\$12\catcode
  `\&12\catcode `\#12\catcode `\^12\catcode `\_12\catcode `\%12\relax}%
\providecommand \@@startlink[1]{}%
\providecommand \@@endlink[0]{}%
\providecommand \url  [0]{\begingroup\@sanitize@url \@url }%
\providecommand \@url [1]{\endgroup\@href {#1}{\urlprefix }}%
\providecommand \urlprefix  [0]{URL }%
\providecommand \Eprint [0]{\href }%
\providecommand \doibase [0]{http://dx.doi.org/}%
\providecommand \selectlanguage [0]{\@gobble}%
\providecommand \bibinfo  [0]{\@secondoftwo}%
\providecommand \bibfield  [0]{\@secondoftwo}%
\providecommand \translation [1]{[#1]}%
\providecommand \BibitemOpen [0]{}%
\providecommand \bibitemStop [0]{}%
\providecommand \bibitemNoStop [0]{.\EOS\space}%
\providecommand \EOS [0]{\spacefactor3000\relax}%
\providecommand \BibitemShut  [1]{\csname bibitem#1\endcsname}%
\let\auto@bib@innerbib\@empty
\bibitem [{\citenamefont {Altman}\ \emph {et~al.}(2003)\citenamefont {Altman},
  \citenamefont {Hofstetter}, \citenamefont {Demler},\ and\ \citenamefont
  {Lukin}}]{altman}%
  \BibitemOpen
  \bibfield  {author} {\bibinfo {author} {\bibfnamefont {E.}~\bibnamefont
  {Altman}}, \bibinfo {author} {\bibfnamefont {W.}~\bibnamefont {Hofstetter}},
  \bibinfo {author} {\bibfnamefont {E.}~\bibnamefont {Demler}}, \ and\ \bibinfo
  {author} {\bibfnamefont {M.~D.}\ \bibnamefont {Lukin}},\ }\href
  {http://stacks.iop.org/1367-2630/5/i=1/a=113} {\bibfield  {journal} {\bibinfo
   {journal} {New Journal of Physics}\ }\textbf {\bibinfo {volume} {5}},\
  \bibinfo {pages} {113} (\bibinfo {year} {2003})}\BibitemShut {NoStop}%
\bibitem [{\citenamefont {Duan}\ \emph {et~al.}(2003)\citenamefont {Duan},
  \citenamefont {Demler},\ and\ \citenamefont {Lukin}}]{duan}%
  \BibitemOpen
  \bibfield  {author} {\bibinfo {author} {\bibfnamefont {L.-M.}\ \bibnamefont
  {Duan}}, \bibinfo {author} {\bibfnamefont {E.}~\bibnamefont {Demler}}, \ and\
  \bibinfo {author} {\bibfnamefont {M.~D.}\ \bibnamefont {Lukin}},\ }\href
  {\doibase 10.1103/PhysRevLett.91.090402} {\bibfield  {journal} {\bibinfo
  {journal} {Phys. Rev. Lett.}\ }\textbf {\bibinfo {volume} {91}},\ \bibinfo
  {pages} {090402} (\bibinfo {year} {2003})}\BibitemShut {NoStop}%
\bibitem [{\citenamefont {Schachenmayer}\ \emph {et~al.}(2015)\citenamefont
  {Schachenmayer}, \citenamefont {Weld}, \citenamefont {Miyake}, \citenamefont
  {Siviloglou}, \citenamefont {Ketterle},\ and\ \citenamefont {Daley}}]{bec4}%
  \BibitemOpen
  \bibfield  {author} {\bibinfo {author} {\bibfnamefont {J.}~\bibnamefont
  {Schachenmayer}}, \bibinfo {author} {\bibfnamefont {D.~M.}\ \bibnamefont
  {Weld}}, \bibinfo {author} {\bibfnamefont {H.}~\bibnamefont {Miyake}},
  \bibinfo {author} {\bibfnamefont {G.~A.}\ \bibnamefont {Siviloglou}},
  \bibinfo {author} {\bibfnamefont {W.}~\bibnamefont {Ketterle}}, \ and\
  \bibinfo {author} {\bibfnamefont {A.~J.}\ \bibnamefont {Daley}},\ }\href
  {\doibase 10.1103/PhysRevA.92.041602} {\bibfield  {journal} {\bibinfo
  {journal} {Phys. Rev. A}\ }\textbf {\bibinfo {volume} {92}},\ \bibinfo
  {pages} {041602} (\bibinfo {year} {2015})}\BibitemShut {NoStop}%
\bibitem [{\citenamefont {Chin}\ \emph {et~al.}(2010)\citenamefont {Chin},
  \citenamefont {Grimm}, \citenamefont {Julienne},\ and\ \citenamefont
  {Tiesinga}}]{chin}%
  \BibitemOpen
  \bibfield  {author} {\bibinfo {author} {\bibfnamefont {C.}~\bibnamefont
  {Chin}}, \bibinfo {author} {\bibfnamefont {R.}~\bibnamefont {Grimm}},
  \bibinfo {author} {\bibfnamefont {P.}~\bibnamefont {Julienne}}, \ and\
  \bibinfo {author} {\bibfnamefont {E.}~\bibnamefont {Tiesinga}},\ }\href
  {\doibase 10.1103/RevModPhys.82.1225} {\bibfield  {journal} {\bibinfo
  {journal} {Rev. Mod. Phys.}\ }\textbf {\bibinfo {volume} {82}},\ \bibinfo
  {pages} {1225} (\bibinfo {year} {2010})}\BibitemShut {NoStop}%
\bibitem [{\citenamefont {Hart}\ \emph {et~al.}(2015)\citenamefont {Hart},
  \citenamefont {Duarte}, \citenamefont {Yang}, \citenamefont {Liu},
  \citenamefont {Paiva}, \citenamefont {Khatami}, \citenamefont {Scalettar},
  \citenamefont {Trivedi}, \citenamefont {Huse},\ and\ \citenamefont
  {Hulet}}]{hart}%
  \BibitemOpen
  \bibfield  {author} {\bibinfo {author} {\bibfnamefont {R.~A.}\ \bibnamefont
  {Hart}}, \bibinfo {author} {\bibfnamefont {P.~M.}\ \bibnamefont {Duarte}},
  \bibinfo {author} {\bibfnamefont {T.-L.}\ \bibnamefont {Yang}}, \bibinfo
  {author} {\bibfnamefont {X.}~\bibnamefont {Liu}}, \bibinfo {author}
  {\bibfnamefont {T.}~\bibnamefont {Paiva}}, \bibinfo {author} {\bibfnamefont
  {E.}~\bibnamefont {Khatami}}, \bibinfo {author} {\bibfnamefont {R.~T.}\
  \bibnamefont {Scalettar}}, \bibinfo {author} {\bibfnamefont {N.}~\bibnamefont
  {Trivedi}}, \bibinfo {author} {\bibfnamefont {D.~A.}\ \bibnamefont {Huse}}, \
  and\ \bibinfo {author} {\bibfnamefont {R.~G.}\ \bibnamefont {Hulet}},\ }\href
  {http://dx.doi.org/10.1038/nature14223} {\bibfield  {journal} {\bibinfo
  {journal} {Nature}\ }\textbf {\bibinfo {volume} {519}},\ \bibinfo {pages}
  {211 EP } (\bibinfo {year} {2015})}\BibitemShut {NoStop}%
\bibitem [{\citenamefont {Mazurenko}\ \emph {et~al.}(2017)\citenamefont
  {Mazurenko}, \citenamefont {Chiu}, \citenamefont {Ji}, \citenamefont
  {Parsons}, \citenamefont {Kan{\'a}sz-Nagy}, \citenamefont {Schmidt},
  \citenamefont {Grusdt}, \citenamefont {Demler}, \citenamefont {Greif},\ and\
  \citenamefont {Greiner}}]{mazurenko}%
  \BibitemOpen
  \bibfield  {author} {\bibinfo {author} {\bibfnamefont {A.}~\bibnamefont
  {Mazurenko}}, \bibinfo {author} {\bibfnamefont {C.~S.}\ \bibnamefont {Chiu}},
  \bibinfo {author} {\bibfnamefont {G.}~\bibnamefont {Ji}}, \bibinfo {author}
  {\bibfnamefont {M.~F.}\ \bibnamefont {Parsons}}, \bibinfo {author}
  {\bibfnamefont {M.}~\bibnamefont {Kan{\'a}sz-Nagy}}, \bibinfo {author}
  {\bibfnamefont {R.}~\bibnamefont {Schmidt}}, \bibinfo {author} {\bibfnamefont
  {F.}~\bibnamefont {Grusdt}}, \bibinfo {author} {\bibfnamefont
  {E.}~\bibnamefont {Demler}}, \bibinfo {author} {\bibfnamefont
  {D.}~\bibnamefont {Greif}}, \ and\ \bibinfo {author} {\bibfnamefont
  {M.}~\bibnamefont {Greiner}},\ }\href {http://dx.doi.org/10.1038/nature22362}
  {\bibfield  {journal} {\bibinfo  {journal} {Nature}\ }\textbf {\bibinfo
  {volume} {545}},\ \bibinfo {pages} {462 EP } (\bibinfo {year}
  {2017})}\BibitemShut {NoStop}%
\bibitem [{\citenamefont {Brown}\ \emph {et~al.}(2017)\citenamefont {Brown},
  \citenamefont {Mitra}, \citenamefont {Guardado-Sanchez}, \citenamefont
  {Schau{\ss}}, \citenamefont {Kondov}, \citenamefont {Khatami}, \citenamefont
  {Paiva}, \citenamefont {Trivedi}, \citenamefont {Huse},\ and\ \citenamefont
  {Bakr}}]{brown}%
  \BibitemOpen
  \bibfield  {author} {\bibinfo {author} {\bibfnamefont {P.~T.}\ \bibnamefont
  {Brown}}, \bibinfo {author} {\bibfnamefont {D.}~\bibnamefont {Mitra}},
  \bibinfo {author} {\bibfnamefont {E.}~\bibnamefont {Guardado-Sanchez}},
  \bibinfo {author} {\bibfnamefont {P.}~\bibnamefont {Schau{\ss}}}, \bibinfo
  {author} {\bibfnamefont {S.~S.}\ \bibnamefont {Kondov}}, \bibinfo {author}
  {\bibfnamefont {E.}~\bibnamefont {Khatami}}, \bibinfo {author} {\bibfnamefont
  {T.}~\bibnamefont {Paiva}}, \bibinfo {author} {\bibfnamefont
  {N.}~\bibnamefont {Trivedi}}, \bibinfo {author} {\bibfnamefont {D.~A.}\
  \bibnamefont {Huse}}, \ and\ \bibinfo {author} {\bibfnamefont {W.~S.}\
  \bibnamefont {Bakr}},\ }\href {\doibase 10.1126/science.aam7838} {\bibfield
  {journal} {\bibinfo  {journal} {Science}\ }\textbf {\bibinfo {volume}
  {357}},\ \bibinfo {pages} {1385} (\bibinfo {year} {2017})}\BibitemShut
  {NoStop}%
\bibitem [{\citenamefont {Cheuk}\ \emph {et~al.}(2016)\citenamefont {Cheuk},
  \citenamefont {Nichols}, \citenamefont {Lawrence}, \citenamefont {Okan},
  \citenamefont {Zhang}, \citenamefont {Khatami}, \citenamefont {Trivedi},
  \citenamefont {Paiva}, \citenamefont {Rigol},\ and\ \citenamefont
  {Zwierlein}}]{cheuk}%
  \BibitemOpen
  \bibfield  {author} {\bibinfo {author} {\bibfnamefont {L.~W.}\ \bibnamefont
  {Cheuk}}, \bibinfo {author} {\bibfnamefont {M.~A.}\ \bibnamefont {Nichols}},
  \bibinfo {author} {\bibfnamefont {K.~R.}\ \bibnamefont {Lawrence}}, \bibinfo
  {author} {\bibfnamefont {M.}~\bibnamefont {Okan}}, \bibinfo {author}
  {\bibfnamefont {H.}~\bibnamefont {Zhang}}, \bibinfo {author} {\bibfnamefont
  {E.}~\bibnamefont {Khatami}}, \bibinfo {author} {\bibfnamefont
  {N.}~\bibnamefont {Trivedi}}, \bibinfo {author} {\bibfnamefont
  {T.}~\bibnamefont {Paiva}}, \bibinfo {author} {\bibfnamefont
  {M.}~\bibnamefont {Rigol}}, \ and\ \bibinfo {author} {\bibfnamefont {M.~W.}\
  \bibnamefont {Zwierlein}},\ }\href {\doibase 10.1126/science.aag3349}
  {\bibfield  {journal} {\bibinfo  {journal} {Science}\ }\textbf {\bibinfo
  {volume} {353}},\ \bibinfo {pages} {1260} (\bibinfo {year}
  {2016})}\BibitemShut {NoStop}%
\bibitem [{\citenamefont {Affleck}(1989)}]{afflek}%
  \BibitemOpen
  \bibfield  {author} {\bibinfo {author} {\bibfnamefont {I.}~\bibnamefont
  {Affleck}},\ }\href {http://stacks.iop.org/0953-8984/1/i=19/a=001} {\bibfield
   {journal} {\bibinfo  {journal} {Journal of Physics: Condensed Matter}\
  }\textbf {\bibinfo {volume} {1}},\ \bibinfo {pages} {3047} (\bibinfo {year}
  {1989})}\BibitemShut {NoStop}%
\bibitem [{\citenamefont {Renard}\ \emph {et~al.}(2003)\citenamefont {Renard},
  \citenamefont {Regnault},\ and\ \citenamefont {Verdaguer}}]{renard}%
  \BibitemOpen
  \bibfield  {author} {\bibinfo {author} {\bibfnamefont {J.-P.}\ \bibnamefont
  {Renard}}, \bibinfo {author} {\bibfnamefont {L.-P.}\ \bibnamefont
  {Regnault}}, \ and\ \bibinfo {author} {\bibfnamefont {M.}~\bibnamefont
  {Verdaguer}},\ }\enquote {\bibinfo {title} {Haldane quantum spin chains},}\
  in\ \href {\doibase 10.1002/9783527620548.ch2} {\emph {\bibinfo {booktitle}
  {Magnetism: Molecules to Materials}}}\ (\bibinfo  {publisher}
  {Wiley-Blackwell},\ \bibinfo {year} {2003})\ Chap.~\bibinfo {chapter} {2},
  pp.\ \bibinfo {pages} {49--93}\BibitemShut {NoStop}%
\bibitem [{\citenamefont {Wierschem}\ and\ \citenamefont
  {Sengupta}(2014)}]{wierschem2}%
  \BibitemOpen
  \bibfield  {author} {\bibinfo {author} {\bibfnamefont {K.}~\bibnamefont
  {Wierschem}}\ and\ \bibinfo {author} {\bibfnamefont {P.}~\bibnamefont
  {Sengupta}},\ }\href {\doibase 10.1103/PhysRevLett.112.247203} {\bibfield
  {journal} {\bibinfo  {journal} {Phys. Rev. Lett.}\ }\textbf {\bibinfo
  {volume} {112}},\ \bibinfo {pages} {247203} (\bibinfo {year}
  {2014})}\BibitemShut {NoStop}%
\bibitem [{\citenamefont {Campbell}\ \emph {et~al.}(2006)\citenamefont
  {Campbell}, \citenamefont {Mun}, \citenamefont {Boyd}, \citenamefont
  {Medley}, \citenamefont {Leanhardt}, \citenamefont {Marcassa}, \citenamefont
  {Pritchard},\ and\ \citenamefont {Ketterle}}]{campbell}%
  \BibitemOpen
  \bibfield  {author} {\bibinfo {author} {\bibfnamefont {G.~K.}\ \bibnamefont
  {Campbell}}, \bibinfo {author} {\bibfnamefont {J.}~\bibnamefont {Mun}},
  \bibinfo {author} {\bibfnamefont {M.}~\bibnamefont {Boyd}}, \bibinfo {author}
  {\bibfnamefont {P.}~\bibnamefont {Medley}}, \bibinfo {author} {\bibfnamefont
  {A.~E.}\ \bibnamefont {Leanhardt}}, \bibinfo {author} {\bibfnamefont {L.~G.}\
  \bibnamefont {Marcassa}}, \bibinfo {author} {\bibfnamefont {D.~E.}\
  \bibnamefont {Pritchard}}, \ and\ \bibinfo {author} {\bibfnamefont
  {W.}~\bibnamefont {Ketterle}},\ }\href {\doibase 10.1126/science.1130365}
  {\bibfield  {journal} {\bibinfo  {journal} {Science}\ }\textbf {\bibinfo
  {volume} {313}},\ \bibinfo {pages} {649} (\bibinfo {year}
  {2006})}\BibitemShut {NoStop}%
\bibitem [{\citenamefont {Dimitrova}\ \emph {et~al.}(2017)\citenamefont
  {Dimitrova}, \citenamefont {Lunden}, \citenamefont {Amato-Grill},
  \citenamefont {Jepsen}, \citenamefont {Yu}, \citenamefont {Messer},
  \citenamefont {Rigaldo}, \citenamefont {Puentes}, \citenamefont {Weld},\ and\
  \citenamefont {Ketterle}}]{dimitrova}%
  \BibitemOpen
  \bibfield  {author} {\bibinfo {author} {\bibfnamefont {I.}~\bibnamefont
  {Dimitrova}}, \bibinfo {author} {\bibfnamefont {W.}~\bibnamefont {Lunden}},
  \bibinfo {author} {\bibfnamefont {J.}~\bibnamefont {Amato-Grill}}, \bibinfo
  {author} {\bibfnamefont {N.}~\bibnamefont {Jepsen}}, \bibinfo {author}
  {\bibfnamefont {Y.}~\bibnamefont {Yu}}, \bibinfo {author} {\bibfnamefont
  {M.}~\bibnamefont {Messer}}, \bibinfo {author} {\bibfnamefont
  {T.}~\bibnamefont {Rigaldo}}, \bibinfo {author} {\bibfnamefont
  {G.}~\bibnamefont {Puentes}}, \bibinfo {author} {\bibfnamefont
  {D.}~\bibnamefont {Weld}}, \ and\ \bibinfo {author} {\bibfnamefont
  {W.}~\bibnamefont {Ketterle}},\ }\href {\doibase 10.1103/PhysRevA.96.051603}
  {\bibfield  {journal} {\bibinfo  {journal} {Phys. Rev. A}\ }\textbf {\bibinfo
  {volume} {96}},\ \bibinfo {pages} {051603} (\bibinfo {year}
  {2017})}\BibitemShut {NoStop}%
\bibitem [{\citenamefont {Bakr}\ \emph {et~al.}(2011)\citenamefont {Bakr},
  \citenamefont {Preiss}, \citenamefont {Tai}, \citenamefont {Ma},
  \citenamefont {Simon},\ and\ \citenamefont {Greiner}}]{bakr}%
  \BibitemOpen
  \bibfield  {author} {\bibinfo {author} {\bibfnamefont {W.~S.}\ \bibnamefont
  {Bakr}}, \bibinfo {author} {\bibfnamefont {P.~M.}\ \bibnamefont {Preiss}},
  \bibinfo {author} {\bibfnamefont {M.~E.}\ \bibnamefont {Tai}}, \bibinfo
  {author} {\bibfnamefont {R.}~\bibnamefont {Ma}}, \bibinfo {author}
  {\bibfnamefont {J.}~\bibnamefont {Simon}}, \ and\ \bibinfo {author}
  {\bibfnamefont {M.}~\bibnamefont {Greiner}},\ }\href
  {http://dx.doi.org/10.1038/nature10668} {\bibfield  {journal} {\bibinfo
  {journal} {Nature}\ }\textbf {\bibinfo {volume} {480}},\ \bibinfo {pages}
  {500 EP } (\bibinfo {year} {2011})}\BibitemShut {NoStop}%
\bibitem [{\citenamefont {Arimondo}\ \emph {et~al.}(1977)\citenamefont
  {Arimondo}, \citenamefont {Inguscio},\ and\ \citenamefont
  {Violino}}]{inguscio}%
  \BibitemOpen
  \bibfield  {author} {\bibinfo {author} {\bibfnamefont {E.}~\bibnamefont
  {Arimondo}}, \bibinfo {author} {\bibfnamefont {M.}~\bibnamefont {Inguscio}},
  \ and\ \bibinfo {author} {\bibfnamefont {P.}~\bibnamefont {Violino}},\ }\href
  {\doibase 10.1103/RevModPhys.49.31} {\bibfield  {journal} {\bibinfo
  {journal} {Rev. Mod. Phys.}\ }\textbf {\bibinfo {volume} {49}},\ \bibinfo
  {pages} {31} (\bibinfo {year} {1977})}\BibitemShut {NoStop}%
\bibitem [{\citenamefont {Gross}\ \emph {et~al.}(2011)\citenamefont {Gross},
  \citenamefont {Shotan}, \citenamefont {Machtey}, \citenamefont {Kokkelmans},\
  and\ \citenamefont {Khaykovich}}]{gross2}%
  \BibitemOpen
  \bibfield  {author} {\bibinfo {author} {\bibfnamefont {N.}~\bibnamefont
  {Gross}}, \bibinfo {author} {\bibfnamefont {Z.}~\bibnamefont {Shotan}},
  \bibinfo {author} {\bibfnamefont {O.}~\bibnamefont {Machtey}}, \bibinfo
  {author} {\bibfnamefont {S.}~\bibnamefont {Kokkelmans}}, \ and\ \bibinfo
  {author} {\bibfnamefont {L.}~\bibnamefont {Khaykovich}},\ }\href {\doibase
  https://doi.org/10.1016/j.crhy.2010.10.004} {\bibfield  {journal} {\bibinfo
  {journal} {Comptes Rendus Physique}\ }\textbf {\bibinfo {volume} {12}},\
  \bibinfo {pages} {4 } (\bibinfo {year} {2011})},\ \bibinfo {note} {few body
  problem}\BibitemShut {NoStop}%
\bibitem [{\citenamefont {Navon}\ \emph {et~al.}(2011)\citenamefont {Navon},
  \citenamefont {Piatecki}, \citenamefont {G\"unter}, \citenamefont {Rem},
  \citenamefont {Nguyen}, \citenamefont {Chevy}, \citenamefont {Krauth},\ and\
  \citenamefont {Salomon}}]{navon}%
  \BibitemOpen
  \bibfield  {author} {\bibinfo {author} {\bibfnamefont {N.}~\bibnamefont
  {Navon}}, \bibinfo {author} {\bibfnamefont {S.}~\bibnamefont {Piatecki}},
  \bibinfo {author} {\bibfnamefont {K.}~\bibnamefont {G\"unter}}, \bibinfo
  {author} {\bibfnamefont {B.}~\bibnamefont {Rem}}, \bibinfo {author}
  {\bibfnamefont {T.~C.}\ \bibnamefont {Nguyen}}, \bibinfo {author}
  {\bibfnamefont {F.}~\bibnamefont {Chevy}}, \bibinfo {author} {\bibfnamefont
  {W.}~\bibnamefont {Krauth}}, \ and\ \bibinfo {author} {\bibfnamefont
  {C.}~\bibnamefont {Salomon}},\ }\href {\doibase
  10.1103/PhysRevLett.107.135301} {\bibfield  {journal} {\bibinfo  {journal}
  {Phys. Rev. Lett.}\ }\textbf {\bibinfo {volume} {107}},\ \bibinfo {pages}
  {135301} (\bibinfo {year} {2011})}\BibitemShut {NoStop}%
\bibitem [{\citenamefont {Pollack}\ \emph
  {et~al.}(2009{\natexlab{a}})\citenamefont {Pollack}, \citenamefont {Dries},
  \citenamefont {Junker}, \citenamefont {Chen}, \citenamefont {Corcovilos},\
  and\ \citenamefont {Hulet}}]{pollack}%
  \BibitemOpen
  \bibfield  {author} {\bibinfo {author} {\bibfnamefont {S.~E.}\ \bibnamefont
  {Pollack}}, \bibinfo {author} {\bibfnamefont {D.}~\bibnamefont {Dries}},
  \bibinfo {author} {\bibfnamefont {M.}~\bibnamefont {Junker}}, \bibinfo
  {author} {\bibfnamefont {Y.~P.}\ \bibnamefont {Chen}}, \bibinfo {author}
  {\bibfnamefont {T.~A.}\ \bibnamefont {Corcovilos}}, \ and\ \bibinfo {author}
  {\bibfnamefont {R.~G.}\ \bibnamefont {Hulet}},\ }\href {\doibase
  10.1103/PhysRevLett.102.090402} {\bibfield  {journal} {\bibinfo  {journal}
  {Phys. Rev. Lett.}\ }\textbf {\bibinfo {volume} {102}},\ \bibinfo {pages}
  {090402} (\bibinfo {year} {2009}{\natexlab{a}})}\BibitemShut {NoStop}%
\bibitem [{\citenamefont {Schori}\ \emph {et~al.}(2004)\citenamefont {Schori},
  \citenamefont {St\"oferle}, \citenamefont {Moritz}, \citenamefont {K\"ohl},\
  and\ \citenamefont {Esslinger}}]{schori}%
  \BibitemOpen
  \bibfield  {author} {\bibinfo {author} {\bibfnamefont {C.}~\bibnamefont
  {Schori}}, \bibinfo {author} {\bibfnamefont {T.}~\bibnamefont {St\"oferle}},
  \bibinfo {author} {\bibfnamefont {H.}~\bibnamefont {Moritz}}, \bibinfo
  {author} {\bibfnamefont {M.}~\bibnamefont {K\"ohl}}, \ and\ \bibinfo {author}
  {\bibfnamefont {T.}~\bibnamefont {Esslinger}},\ }\href {\doibase
  10.1103/PhysRevLett.93.240402} {\bibfield  {journal} {\bibinfo  {journal}
  {Phys. Rev. Lett.}\ }\textbf {\bibinfo {volume} {93}},\ \bibinfo {pages}
  {240402} (\bibinfo {year} {2004})}\BibitemShut {NoStop}%
\bibitem [{\citenamefont {Mark}\ \emph {et~al.}(2012)\citenamefont {Mark},
  \citenamefont {Haller}, \citenamefont {Lauber}, \citenamefont {Danzl},
  \citenamefont {Janisch}, \citenamefont {B\"uchler}, \citenamefont {Daley},\
  and\ \citenamefont {N\"agerl}}]{mark}%
  \BibitemOpen
  \bibfield  {author} {\bibinfo {author} {\bibfnamefont {M.~J.}\ \bibnamefont
  {Mark}}, \bibinfo {author} {\bibfnamefont {E.}~\bibnamefont {Haller}},
  \bibinfo {author} {\bibfnamefont {K.}~\bibnamefont {Lauber}}, \bibinfo
  {author} {\bibfnamefont {J.~G.}\ \bibnamefont {Danzl}}, \bibinfo {author}
  {\bibfnamefont {A.}~\bibnamefont {Janisch}}, \bibinfo {author} {\bibfnamefont
  {H.~P.}\ \bibnamefont {B\"uchler}}, \bibinfo {author} {\bibfnamefont {A.~J.}\
  \bibnamefont {Daley}}, \ and\ \bibinfo {author} {\bibfnamefont {H.-C.}\
  \bibnamefont {N\"agerl}},\ }\href {\doibase 10.1103/PhysRevLett.108.215302}
  {\bibfield  {journal} {\bibinfo  {journal} {Phys. Rev. Lett.}\ }\textbf
  {\bibinfo {volume} {108}},\ \bibinfo {pages} {215302} (\bibinfo {year}
  {2012})}\BibitemShut {NoStop}%
\bibitem [{\citenamefont {B\"uchler}(2010)}]{buchler}%
  \BibitemOpen
  \bibfield  {author} {\bibinfo {author} {\bibfnamefont {H.~P.}\ \bibnamefont
  {B\"uchler}},\ }\href {\doibase 10.1103/PhysRevLett.104.090402} {\bibfield
  {journal} {\bibinfo  {journal} {Phys. Rev. Lett.}\ }\textbf {\bibinfo
  {volume} {104}},\ \bibinfo {pages} {090402} (\bibinfo {year}
  {2010})}\BibitemShut {NoStop}%
\bibitem [{\citenamefont {Pollack}\ \emph
  {et~al.}(2009{\natexlab{b}})\citenamefont {Pollack}, \citenamefont {Dries},\
  and\ \citenamefont {Hulet}}]{pollack2}%
  \BibitemOpen
  \bibfield  {author} {\bibinfo {author} {\bibfnamefont {S.~E.}\ \bibnamefont
  {Pollack}}, \bibinfo {author} {\bibfnamefont {D.}~\bibnamefont {Dries}}, \
  and\ \bibinfo {author} {\bibfnamefont {R.~G.}\ \bibnamefont {Hulet}},\ }\href
  {\doibase 10.1126/science.1182840} {\bibfield  {journal} {\bibinfo  {journal}
  {Science}\ }\textbf {\bibinfo {volume} {326}},\ \bibinfo {pages} {1683}
  (\bibinfo {year} {2009}{\natexlab{b}})}\BibitemShut {NoStop}%
\bibitem [{\citenamefont {Gross}\ and\ \citenamefont
  {Khaykovich}(2008)}]{gross}%
  \BibitemOpen
  \bibfield  {author} {\bibinfo {author} {\bibfnamefont {N.}~\bibnamefont
  {Gross}}\ and\ \bibinfo {author} {\bibfnamefont {L.}~\bibnamefont
  {Khaykovich}},\ }\href {\doibase 10.1103/PhysRevA.77.023604} {\bibfield
  {journal} {\bibinfo  {journal} {Phys. Rev. A}\ }\textbf {\bibinfo {volume}
  {77}},\ \bibinfo {pages} {023604} (\bibinfo {year} {2008})}\BibitemShut
  {NoStop}%
\bibitem [{\citenamefont {Shotan}\ \emph {et~al.}(2014)\citenamefont {Shotan},
  \citenamefont {Machtey}, \citenamefont {Kokkelmans},\ and\ \citenamefont
  {Khaykovich}}]{shotan}%
  \BibitemOpen
  \bibfield  {author} {\bibinfo {author} {\bibfnamefont {Z.}~\bibnamefont
  {Shotan}}, \bibinfo {author} {\bibfnamefont {O.}~\bibnamefont {Machtey}},
  \bibinfo {author} {\bibfnamefont {S.}~\bibnamefont {Kokkelmans}}, \ and\
  \bibinfo {author} {\bibfnamefont {L.}~\bibnamefont {Khaykovich}},\ }\href
  {\doibase 10.1103/PhysRevLett.113.053202} {\bibfield  {journal} {\bibinfo
  {journal} {Phys. Rev. Lett.}\ }\textbf {\bibinfo {volume} {113}},\ \bibinfo
  {pages} {053202} (\bibinfo {year} {2014})}\BibitemShut {NoStop}%
\end{thebibliography}%

\end{document}